\documentclass[aps,prl,superscriptaddress,twocolumn,longbibliography]{revtex4-1}
\usepackage{bbm}
\usepackage{graphicx}
\usepackage{dcolumn}
\usepackage{bm}
\usepackage{subfigure}
\usepackage{amsmath}
\usepackage{feynmf}
\usepackage{hyperref}

\usepackage{attachfile}

\newcommand{\bk}{\boldsymbol k}

\newcommand{\br}{\boldsymbol r}

\usepackage{times}

\begin{document}
\title{ Vortex end Majorana zero modes in superconducting Dirac and Weyl semimetals}
\author{Zhongbo Yan}
\email{yanzhb5@mail.sysu.edu.cn}
\affiliation{ School of Physics, Sun Yat-Sen University, Guangzhou 510275, China}

\author{Zhigang Wu}
\affiliation{Guangdong Provincial Key Laboratory of Quantum Science and Engineering, Shenzhen Institute for Quantum Science and Engineering, Southern University of Science and Technology, Shenzhen 518055, Guangdong, China}
\affiliation{Center for Quantum Computing, Peng Cheng Laboratory, Shenzhen 518005, China}

\author{Wen Huang}
\affiliation{Guangdong Provincial Key Laboratory of Quantum Science and Engineering, Shenzhen Institute for Quantum Science and Engineering, Southern University of Science and Technology, Shenzhen 518055, Guangdong, China}
\date{\today}

\begin{abstract}

Time-reversal invariant (TRI) Dirac and Weyl semimetals in three dimensions (3D) can host
open Fermi arcs and spin-momentum locking Fermi loops on the surfaces. We find
that when they become superconducting with $s$-wave pairing and the doping is lower than
a critical level, straight $\pi$-flux vortex lines terminating at surfaces with Fermi arcs or spin-momentum locking Fermi loops can realize 1D topological superconductivity and harbor Majorana
zero modes at their ends. Remarkably, we find that the vortex-generation-associated Zeeman field can open (when the surfaces have
only Fermi arcs) or enhance the topological gap protecting Majorana zero modes, which is contrary to
the situation in superconducting topological insulators. By studying the tilting effect
of bulk Dirac and Weyl cones, we further find that type-I Dirac and Weyl semimetals in general have
a much broader topological regime than type-II ones.
Our findings build up a connection between TRI Dirac and Weyl semimetals and Majorana zero modes in vortices.

\end{abstract}

\maketitle

Majorana zero modes (MZMs) localised in the vortices of the superconducting order parameter have been actively sought after for more than a
decade\cite{alicea2012new,Beenakker2013,stanescu2013majorana, leijnse2012introduction,Elliott2015,sarma2015majorana,sato2016majorana,aguado2017}. Originally, they were predicted to exist in 2D chiral $p$-wave superconductors\cite{read2000},
but the scarcity of such superconductors in nature compels the community to look for
other possible candidates. A breakthrough comes with the realization that
when the spin-momentum locking Dirac surface states of 3D topological insulators (TIs)
are gapped by $s$-wave superconductivity, a $\pi$-flux vortex carries a
single MZM at its core\cite{fu2008}.  Since then,
TIs with either proximity-induced
or  intrinsic  superconductivity are strongly desired for the realization of
MZMs in experiment\cite{Hor2010,wray2010observation,Kriener2011,Sasaki2011,wang2012coexistence, Wang2013proximity,Zareapour2012proximity,Xu2015MZM,lv2016,Sun2016Majorana}.
Remarkably,  the normal state of several iron-based  superconductors has recently been found to
carry an inverted band structure and thus Dirac surface states\cite{Wang2015iFeSC,Xu2016FeSC,zhang2019multiple}.
In superconducting states, zero-bias peaks, as well as several other ordered discrete
peaks, are clearly observed in scanning tunnelling spectroscopy of these materials when the tip moves close to the vortex core\cite{zhang2018iron,wang2018evidence,Liu2018MZM,machida2019zero,chen2019quantized,Zhu2019MZM,kong2019half}, thus providing
strong experimental evidences for the existence of MZMs in the vortices of these superconductors.

In efforts to broaden the scope of materials that can host MZMs in vortices, the question of whether the bulk material needs to be insulating has been naturally raised\cite{Hosur2011MZM,Chiu2011vortex,Hung2013vortex,Qin2019vortex2,Ghazaryan2019vortex,Ghorashi2019vortex}. The finding that the MZMs survive even in the doped TI with metallic normal states\cite{Hosur2011MZM} has motivated us to examine the possibility of MZMs appearing in the recently discovered 3D Dirac and Weyl semimetals\cite{wan2011,Xu2011weyl,wang2012dirac,young2012dirac,wang2013three,Weng2015weyl,huang2015weyl,
Liu2014discovery,neupane2014observation,Borisenko2014,Xu2015science,Lv2015weyl,lu2015}, though their
bulk and boundary physics display remarkable difference compared with TIs\cite{Armitage2018RMP,Yan2016review}. With respect to the bulk,  the energy bands of Dirac and Weyl semimetals
touch at some isolated points (the so-called
Dirac and Weyl points) away from which the dispersions are distinct from those of conventional
quasiparticles in metals and are responsible for some novel transport properties\cite{burkov2015chiral,Wang2017review,Hu2019review}.
On the boundary, a hallmark is the existence of open Fermi arcs
which connect the projections of the bulk Dirac and Weyl points\cite{Xu2015science,Lv2015weyl}. Such Fermi arcs
have attracted great research interest as they can result in many remarkable phenomena\cite{potter2014quantum,moll2016transport,Wang2017Hall,zhang2019quantum}.

Intriguingly,  when time-reversal symmetry is preserved, spin-momentum locking Fermi loops (abbreviated as
``Fermi loops'' below), a manifestation of Dirac surface states, can coexist with the Fermi arcs on the surfaces of
Dirac and Weyl semimetals\cite{Kargarian2016,Lau2017coexist,Le2018coexist,Wu2019coexist}.
With this observation and the scenario in doped TIs in mind,
we study $\pi$-flux vortex lines terminating at such surfaces
and do find the existence of MZMs at the vortex ends below a critical doping level.
Strikingly, we find that the topological gap of
the vortex line protecting MZMs in superconducting Dirac and
Weyl semimetals can be profoundly enhanced by the
vortex-generation-associated Zeeman field, which is contrary to
the situation in superconducting TIs. Importantly,
we also find that the Zeeman field can lead to the realization
of vortex-end MZMs even when the surfaces have only
Fermi arcs, indicating the generality of the underlying physics.
In addition, we find that the vortex lines in type-I Dirac and Weyl semimetals
have a much broader topological regime than those in the type-II ones.

\begin{figure}
\subfigure{\includegraphics[width=4.25cm, height=4cm]{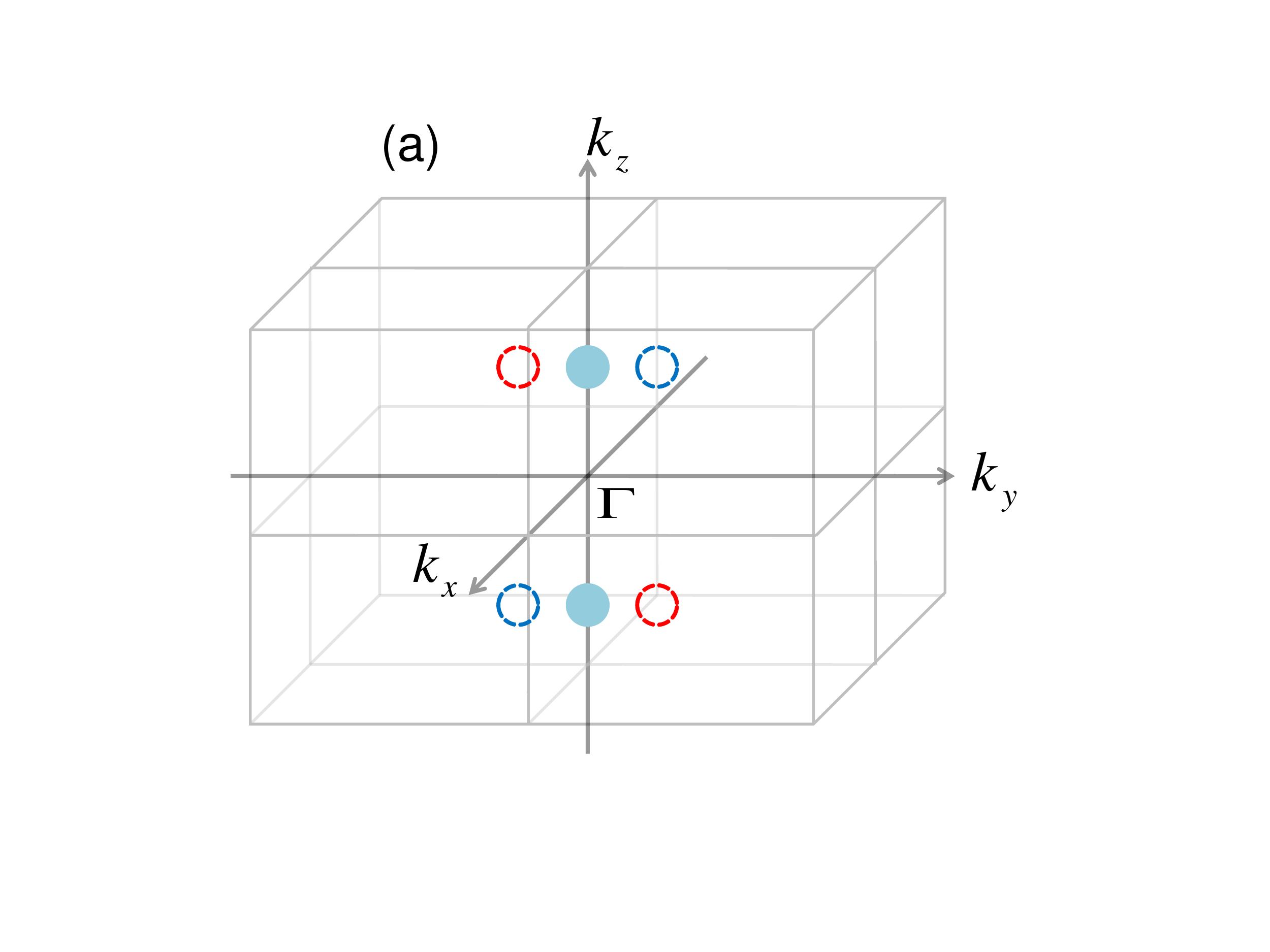}}
\subfigure{\includegraphics[width=4.25cm, height=4cm]{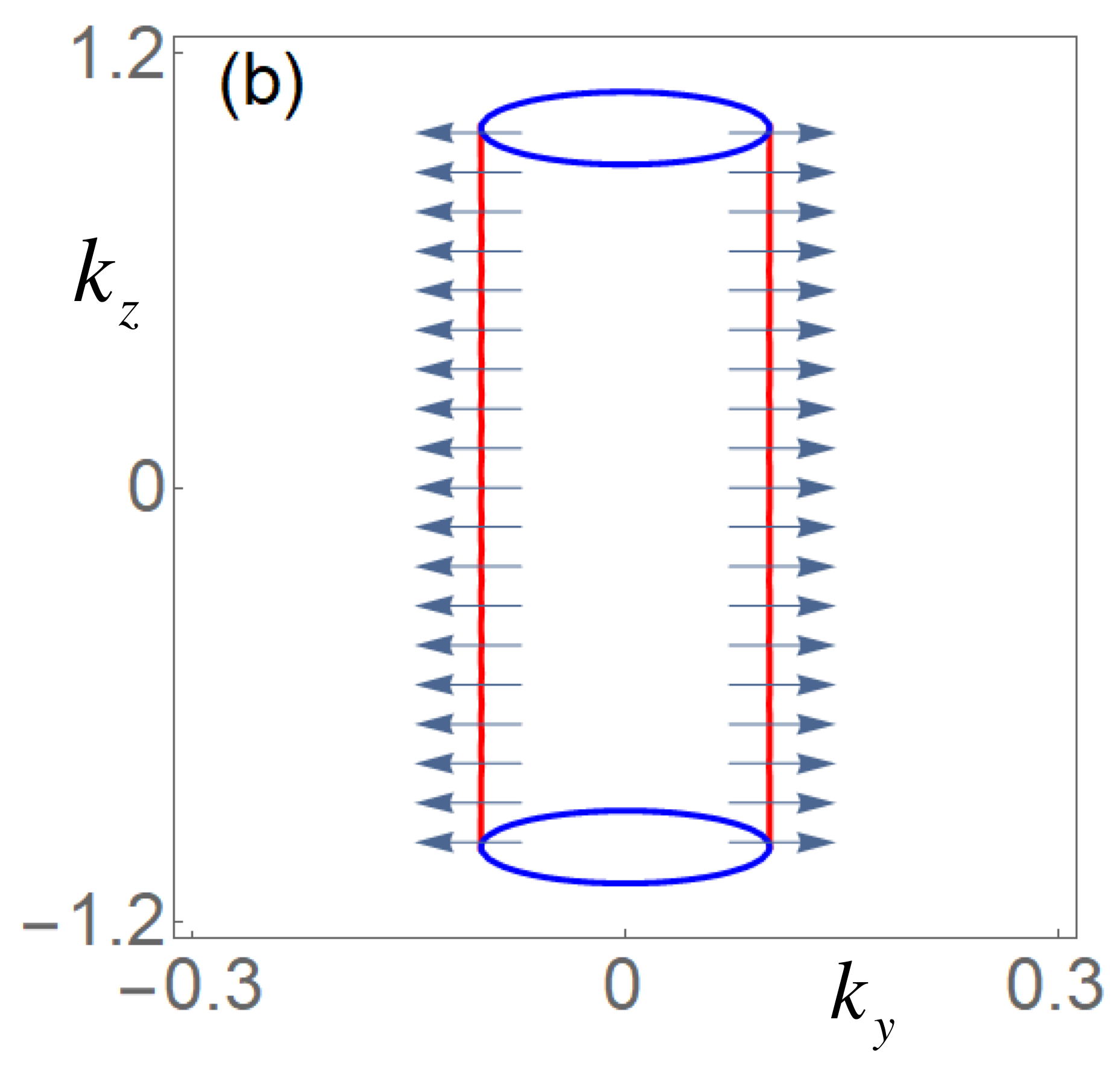}}
\subfigure{\includegraphics[width=4.25cm, height=4cm]{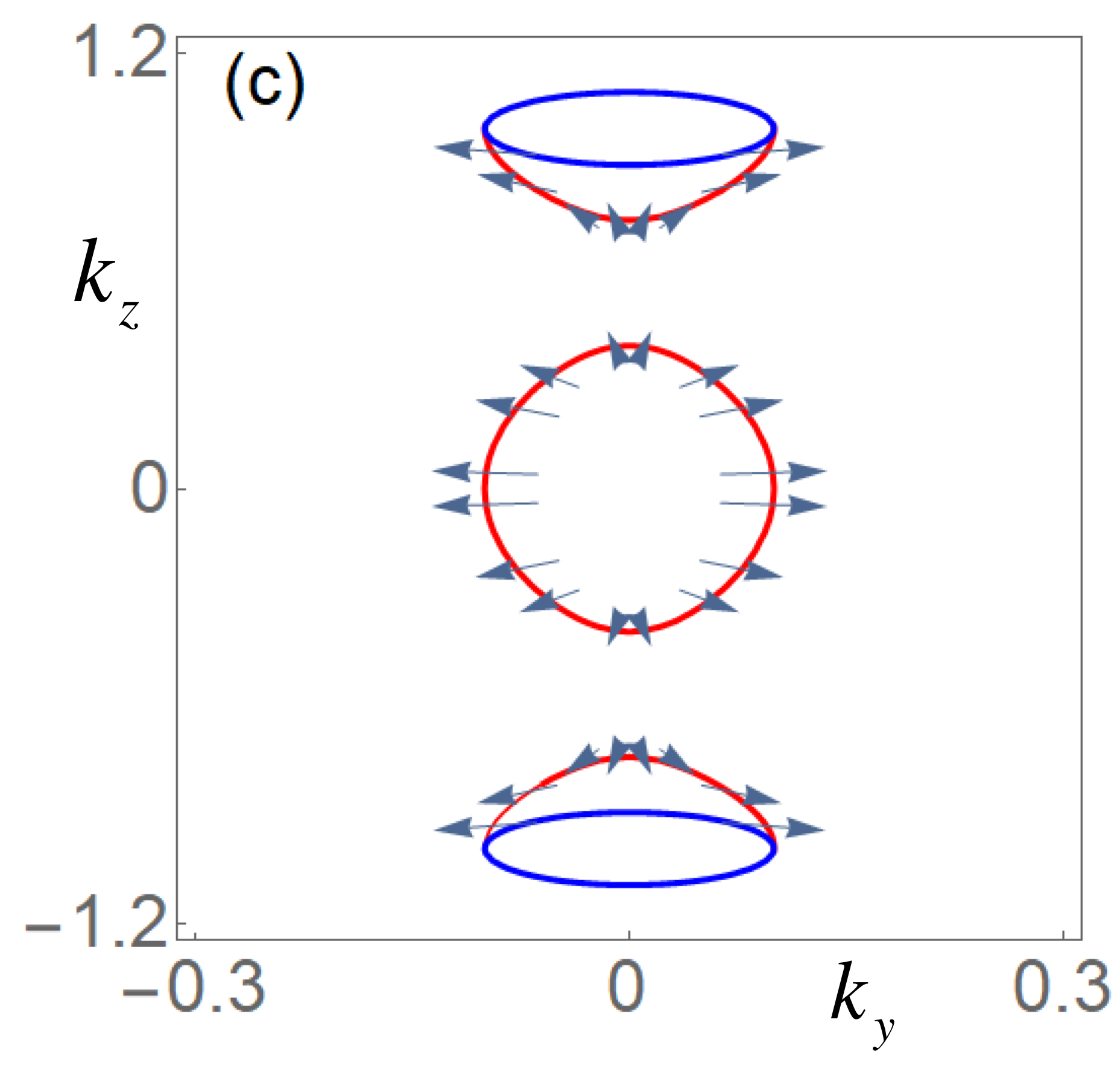}}
\subfigure{\includegraphics[width=4.25cm, height=4cm]{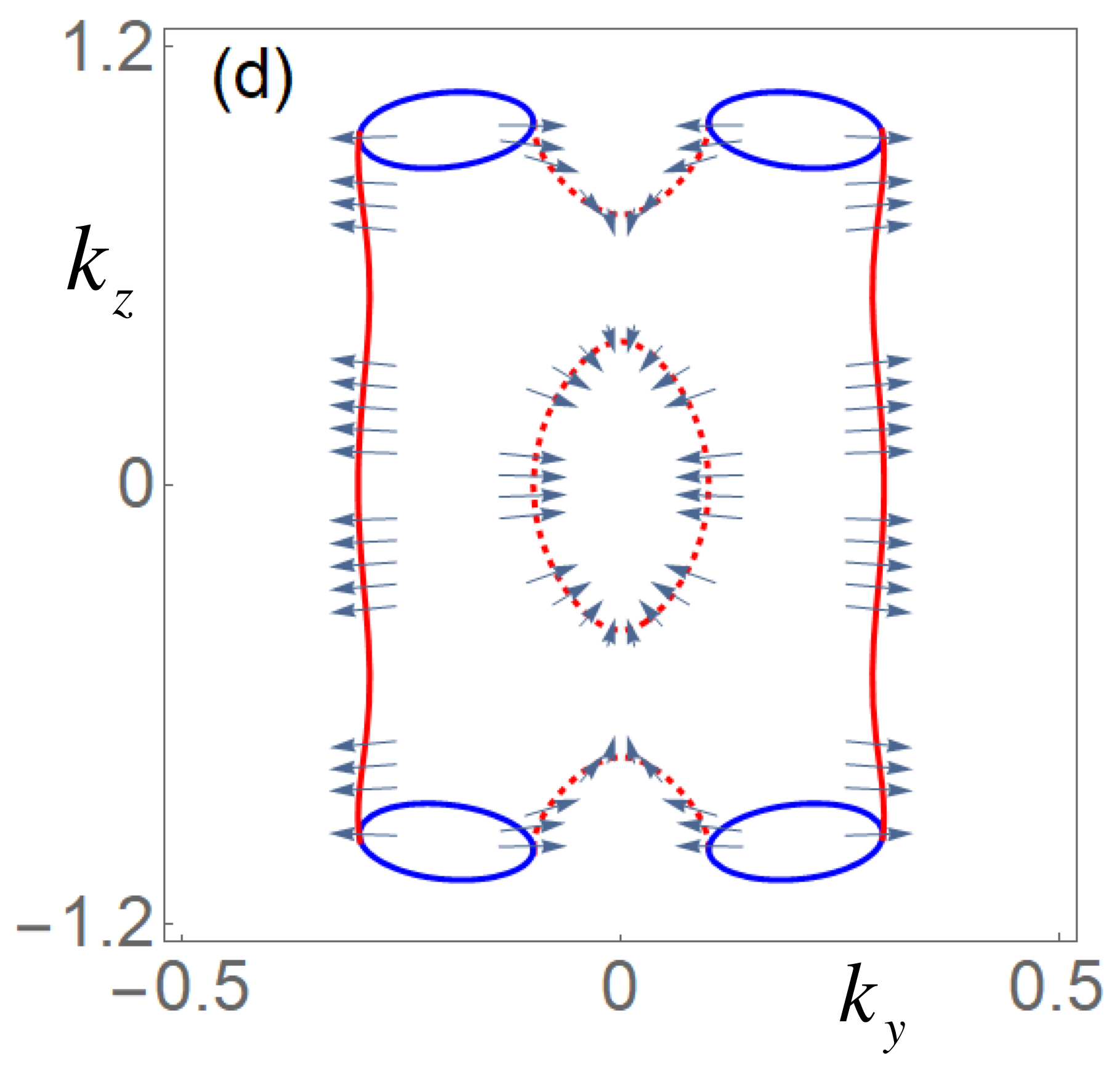}}
\caption{(a) A schematic diagram of Dirac and
Weyl points in Brillouin zone. The filled dots on the $k_{z}$ axis
represent Dirac points. Breaking inversion symmetry splits
the two Dirac points into four Weyl points (the four unfilled dots,
the red and blue colors represent opposite chirality). In (b)(c)(d),
the red lines represent the
constant-energy contours of surface states at energy $E_{s}=0.1$,
the blue lines represent the projection of bulk
Fermi surfaces at Fermi energy $E_{F}=E_{s}$ in the $k_{y}$-$k_{z}$ plane, and the arrows
show the spin textures of the surface states.
Common parameters are $m=2.5$ and $t=t_{3}=t'=1$.
(b) $t_{3}'=\delta=0$. Two open Fermi arcs connect the
projections of bulk Fermi surfaces tangentially.
(c) $t_{3}'=0.6$, $\delta=0$. A  Fermi loop coexists with two Fermi arcs.
(d) $t_{3}'=0.6$, $\delta=0.2$.
The red solid lines are Fermi arcs
on the $x=0$ surface, and the red dashed lines are Fermi arcs and Fermi loop on the $x=L$ surface. }  \label{sketch}
\end{figure}

{\it Fermi arcs and Fermi loops.---} We begin with the normal state Hamiltonian, which reads\cite{Qin2019vortex}
\begin{eqnarray}
H_{0}(\bk)&=&\varepsilon(\bk)+(m-t\cos k_{x}-t\cos k_{y}-t_{3}\cos k_{z})\sigma_{z}\nonumber\\
&&+t'\sin k_{x}\sigma_{x}s_{z}+t_{3}'\sin k_{z}(\cos k_{x}-\cos k_{y})\sigma_{x}s_{x}\nonumber\\
&&-t'\sin k_{y}\sigma_{y}+2t_{3}'\sin k_{z}\sin k_{x}\sin k_{y}\sigma_{x}s_{y}+\delta \sigma_{y}s_{z}\quad\label{normal}
\end{eqnarray}
in the basis $\Psi_{\bk}^{\dag}=(c_{1,\uparrow,\bk}^{\dag}
,c_{2,\uparrow,\bk}^{\dag},c_{1,\downarrow,\bk}^{\dag},c_{2,\downarrow,\bk}^{\dag})$,
where the Pauli matrices $\sigma_{i}$ and $s_{i}$ with $i=\{x,y,z\}$ act in the orbit  and spin spaces respectively. For notational simplicity, identity matrices $\sigma_{0}$ and $s_{0}$ are made implicit and the lattice constant is set to unit throughout the whole work. $\varepsilon(\bk)$ is
a term characterizing the asymmetry of conduction and
valence bands. In this work, we take $\varepsilon(\bk)=0$ for
simplicity unless otherwise specified.
When $\delta=0$, $H_{0}(\bk)$ has time-reversal symmetry, inversion symmetry and C$_{4z}$-rotational symmetry.
These symmetries stabilize Dirac points on rotational-symmetry axes.
A finite $\delta$ breaks the inversion symmetry and consequently splits a Dirac point
into two Weyl points with opposite chirality, as illustrated
in Fig.\ref{sketch}(a).

While the two $t'_{3}$-terms are often ignored in the study of Dirac and Weyl semimetals,
they are in fact ubiquitous in real materials as they are allowed by symmetry\cite{Kargarian2016,Le2018coexist}.
Although they do not (or weakly) affect
the linear dispersion near the Dirac (or Weyl) points, they do have a strong impact on the boundary modes. To see
this, we examine the system occupying
the region $0\leq x \leq L$ ($L$ assumed to be very large) and use the edge theory to
obtain the low-energy Hamiltonians describing the boundary modes on the $x=0$ and $x=L$ surfaces\cite{Yan2018hosc}.
By considering that the band inversion occurs at the $\Gamma$ point,
 we find that they are respectively given by (derivation details
are provided in the Supplemental Material (SM)\cite{supplemental})
\begin{eqnarray}
H_{\rm S;0}(k_{y},k_{z})&=&-\delta+v_{y} k_{y}s_{z}+v_{z}(k_{y},k_{z})k_{z}s_{y},\nonumber\\
H_{\rm S;L}(k_{y},k_{z})&=&\delta-v_{y} k_{y}s_{z}-v_{z}(k_{y},k_{z})k_{z}s_{y},\label{diracsurface}
\label{surface1}
\end{eqnarray}
where  $v_{y}=t'$, $v_{z}(k_{y},k_{z})=-t_{3}'(\tilde{m}+tk_{y}^{2}+t_{3}k_{z}^{2}/2)/t$, $\tilde{m}=m-2t-t_{3}$, 
and the boundary modes exist within the regime
$tk_{y}^{2}/2+t_{3}k_{z}^{2}/2<-\tilde{m}$. In the absence of the $t_{3}'$-terms, i.e., $t_{3}'=0$,
it is readily found that $v_{z}(k_{y},k_{z})=0$ and  each Hamiltonian in Eq.(\ref{surface1}) describes a
pair of 1D helical modes reminiscent of 2D TIs. However, once $t_{3}'\neq0$, each Hamiltonian
in Eq.(\ref{surface1}) describes a 2D Dirac cone at $\bar{\Gamma}=(0,0)$ in the leading order, which is reminiscent of 3D TIs. Moreover, one can see that the $\delta$ term only induces an energy shift of the Dirac cones.
The above analyses show that Dirac surface states
exist not only in TIs, but also in TRI Dirac and Weyl semimetals.

In Figs.\ref{sketch}(b)(c)(d), we present the constant-energy contours of surface states (CECSS) for various choices of $t_{3}'$ and $\delta$.
According to Eq.(\ref{diracsurface}), the CECSS at energy $E_{s}$ are determined by
$\pm\delta\pm \sqrt{v_{y}^{2} k_{y}^{2}+v_{z}^{2}(k_{y},k_{z})k_{z}^{2}}=E_{s}$ under the constraint
$tk_{y}^{2}/2+t_{3}k_{z}^{2}/2<-\tilde{m}$.  Fig.\ref{sketch}(b) shows that when $t_{3}'=\delta=0$,
the CECSS on the $x=0$ surface are two open Fermi arcs whose spin textures are opposite due to time-reversal symmetry.
 It is worth pointing out
that for the Dirac semimetal, i.e., $\delta=0$,  Eq.(\ref{diracsurface}) indicates that the CECSS
on the $x=0$ and $x=L$ surface coincide and display opposite spin textures,
so we only present the results on the $x=0$ surface for clarity. Fig.\ref{sketch}(c)
shows that when $t_{3}'\neq0$, $\delta=0$, the CECSS on the $x=0$ surface
contains two open Fermi arcs and one  Fermi loop. As one Dirac point is equivalent to two
Weyl points with opposite chirality, one can see that the open Fermi arcs in Fig.\ref{sketch}(c) begin and end
at the same projection of bulk Fermi surfaces. When $\delta$ becomes nonzero, the CECSS on the $x=0$ and $x=L$ surfaces no longer coincide. Fig.\ref{sketch}(d) shows that while two Fermi arcs
and one Fermi loop coexist on the $x=L$ surface, the $x=0$ surface only contains two Fermi arcs.

{\it Vortex bound states in superconducting Dirac and Weyl semimetals.---}  Two
groups have recently found that a $\pi$-flux vortex line parallel to
the rotational-symmetry axis
in superconducting Dirac semimetals hosts 1D propagating Majorana
modes\cite{Qin2019vortex,Konig2019votex}. There the vortex line is
terminated at the $z$-normal surfaces without any Fermi arc or loop.
In this work, we consider $\pi$-flux vortex lines along the $x$ direction
and explore the consequence of Fermi arcs and loops on the $x$-normal surfaces.

When the TRI Dirac and Weyl semimetals become superconducting and vortex lines are generated
along the $x$ direction by an external magnetic field, the system is described by
$H=\frac{1}{2}\sum_{\br,\br'}\Psi_{\br}^{\dag}H_{\rm BdG}(\br,\br')\Psi_{\br'}$,
with the basis $\Psi_{\br}^{\dag}=(c_{1,\uparrow,\br}^{\dag}
,c_{2,\uparrow,\br}^{\dag},c_{1,\downarrow,\br}^{\dag},c_{2,\downarrow,\br}^{\dag},c_{1,\uparrow,\br}
,c_{2,\uparrow,\br},c_{1,\downarrow,\br},c_{2,\downarrow,\br})$ and the real-space
Bogoliubov-de Gennes (BdG) Hamiltonian
\begin{eqnarray}
H_{\rm BdG}(\br,\br')=\left(
                   \begin{array}{cc}
                     H_{\br,\br'} & \Delta_{\br,\br'} \\
                     \Delta^{\dag}_{\br',\br} & -H_{\br',\br}^{T} \\
                   \end{array}
                 \right),
\end{eqnarray}
where $H_{\br,\br'}=H_{0;\br,\br'}+V_{z;\br,\br'}-\mu_{\br,\br'}$ with
$H_{0;\br,\br'}$ being the Fourier transformation of
$H_{0}(\bk)$ in Eq.(\ref{normal}), $V_{z;\br,\br'}=\frac{1}{2}g\mu_{B}Bs_{x}\delta_{\br,\br'}\equiv V_{z}s_{x}\delta_{\br,\br'}$ represents the
vortex-generation-associated Zeeman field whose impact depends on both the magnetic field
$B$ and the material-dependent $g$ factor\cite{supplemental}, and $\mu_{\br,\br'}=\mu\delta_{\br,\br'}$
is the chemical potential; $\delta_{\br,\br'}$ is
the Kronecker symbol; $\Delta_{\br,\br'}$ is the superconducting order parameter.
Similarly to Refs.\cite{Qin2019vortex,Konig2019votex}, here we consider
a spin-singlet $s$-wave pairing, i.e., $\Delta_{\br,\br'}=-i\Delta(\br)s_{y}\delta_{\br,\br'}$,
which fully gaps the bulk states when the Zeeman field is smaller than its amplitude.
To model a vortex line in the $x$ direction, we assume  $\Delta(\br)=\Delta_{0}\tanh(\sqrt{y^{2}+z^{2}}/\xi)e^{i\theta}$ where
$\theta=\arctan (y/z)$ and $\xi$ is the coherence length. The vortex line breaks the translational symmetry in the  $yz$ plane and the time-reversal symmetry,  therefore, the 3D superconducting system can be viewed as a 1D superconductor
whose unit cell contains all lattice sites
in the $yz$ plane. Accordingly, the system belongs to the class $D$ of the Atland-Zirnbauer classification
and is characterized by a $Z_{2}$ invariant if its energy spectrum is fully-gapped\cite{schnyder2008,kitaev2009}.  The
$Z_{2}$ invariant is given by\cite{Kitaev2001unpaired,Wimmer2012pfaffian,supplemental}
\begin{eqnarray}
\nu=\text{sgn}(\text{Pf}(A(k_{x}=0)))\cdot\text{sgn}(\text{Pf}(A(k_{x}=\pi))),\label{invariant}
\end{eqnarray}
where $A(k_{x})$ denotes the BdG Hamiltonian in the
Majorana representation i.e., $H=i\sum_{k_{x}}\gamma_{k_{x}}A(k_{x})\gamma_{-k_{x}}$
with $\gamma_{k_{x}}$ denoting the Majorana basis\cite{supplemental}, and
$\text{Pf}(A(k_{x}))$ denotes the Pfaffian of the antisymmetric matrix $A(k_{x})$.
The sign function $\text{sgn}(x)=1(-1)$ for $x>(<)0$. And
$\nu=-1/1$ corresponds to a topological/trivial vortex line with/without
robust MZMs appearing at its ends.

\begin{figure}
\subfigure{\includegraphics[width=8cm, height=6cm]{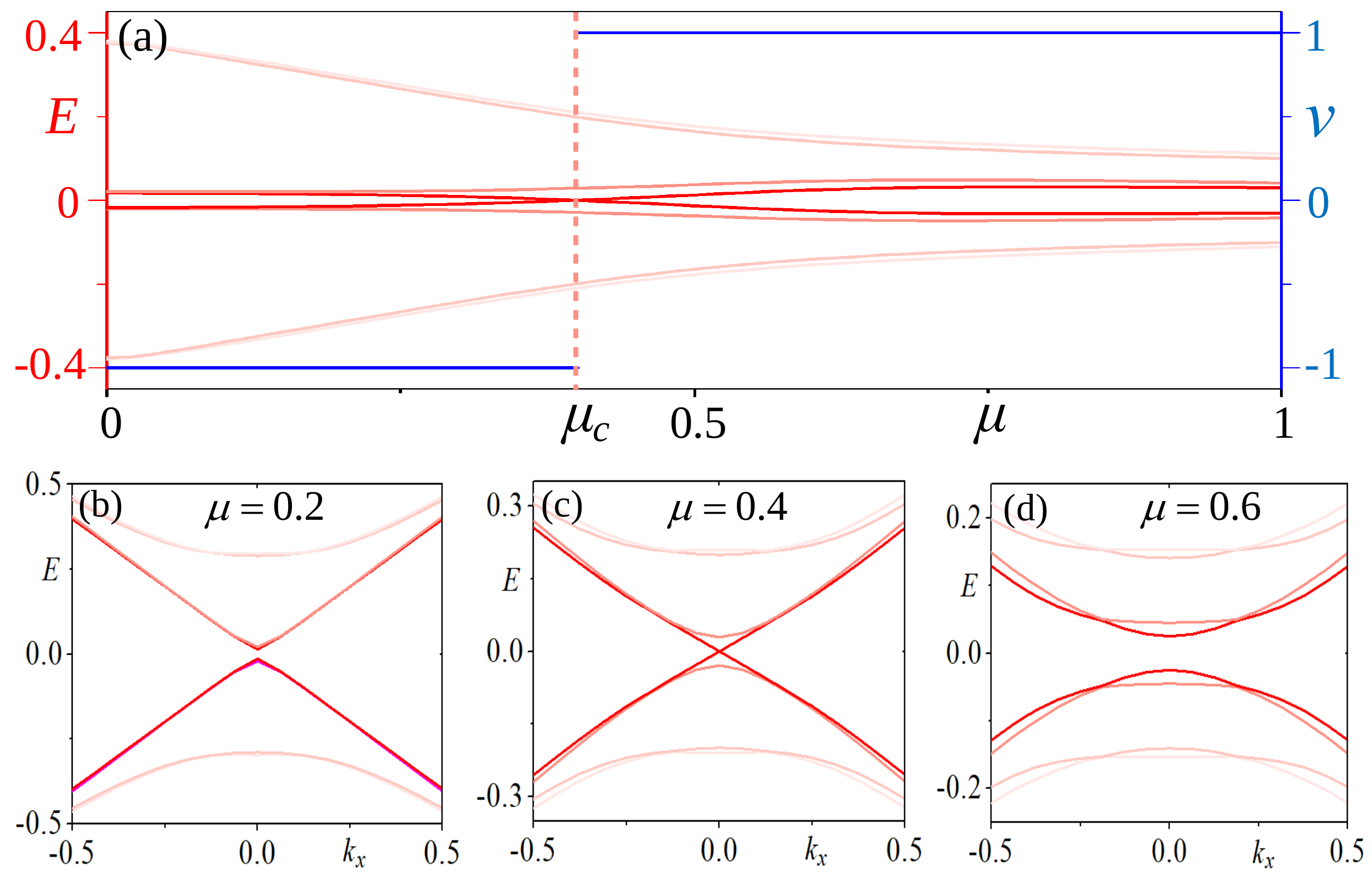}}
\caption{The evolution of vortex bound states (only the lowest few states are shown)
with respect to doping level $\mu$. Common parameters are
$m=2.5$, $t=t_{3}=t'=t_{3}'=1$, $V_{z}=0$, $\delta=0$, $\Delta_{0}=0.5$, $\xi=4$, $L_{y}=L_{z}=20$. (a) Vortex bound states at $k_{x}=0$. There is a crossing
at $\mu_{c}=0.4$, where the $Z_{2}$ invariant (blue line) also jumps and indicates that a vortex phase transition takes place. (b)(c)(d) The dispersions
of vortex bound states at different values of $\mu$. The bound states at $k_{x}=\pi$ remains gapped with the increase of $\mu$, while the bound states at $k_{x}=0$
undergoes a gap closure at $\mu_{c}$.  }  \label{vortex}
\end{figure}

Let us first focus on the superconducting Dirac semimetal with finite $t_{3}'$ terms
and ignore the Zeeman splitting. Because the band inversion is considered to occur at the $\Gamma$ point and
the MZMs are related to the low-energy excitations of the normal state, Eq.(\ref{invariant}) implies
that we can focus on $k_{x}=0$ as long as the Fermi level is not doped too far away from the Dirac points.
In Fig.\ref{vortex}, we present the numerical results of the evolution of the dispersions for vortex bound states
and the $Z_{2}$ invariant under the variation of $\mu$, i.e., the doping level (only $\mu>0$ is shown as the result
is found to be symmetric about
$\mu=0$).
For clarity, we have only shown eight bound states closest to zero energy.
The results reveal that the vortex line is full-gapped except at $\mu_{c}$
where a vortex phase transition takes place.
Similarly to superconducting
TIs\cite{Hosur2011MZM}, we find $\nu=-1$ when the doping is below a critical level,
signaling the realization of 1D
topological superconductivity on the vortex line. In this regime,  each end of the vortex line
binds one MZM. In spite of the similarity
in the topological phase diagram, there are profound distinctions between the vortex lines of superconducting
Dirac semimetals and those of superconducting TIs.
One is that the $\pi$-Berry phase developed to explain the
vortex phase transition for the latter\cite{Hosur2011MZM} does not apply to the former\cite{supplemental},
owing to their fundamental differences in normal states.  Moreover,
as  will be shown below, the effect of Zeeman field to the vortex lines
is completely different for these two classes of systems.

\begin{figure}
\subfigure{\includegraphics[width=4.25cm, height=3.5cm]{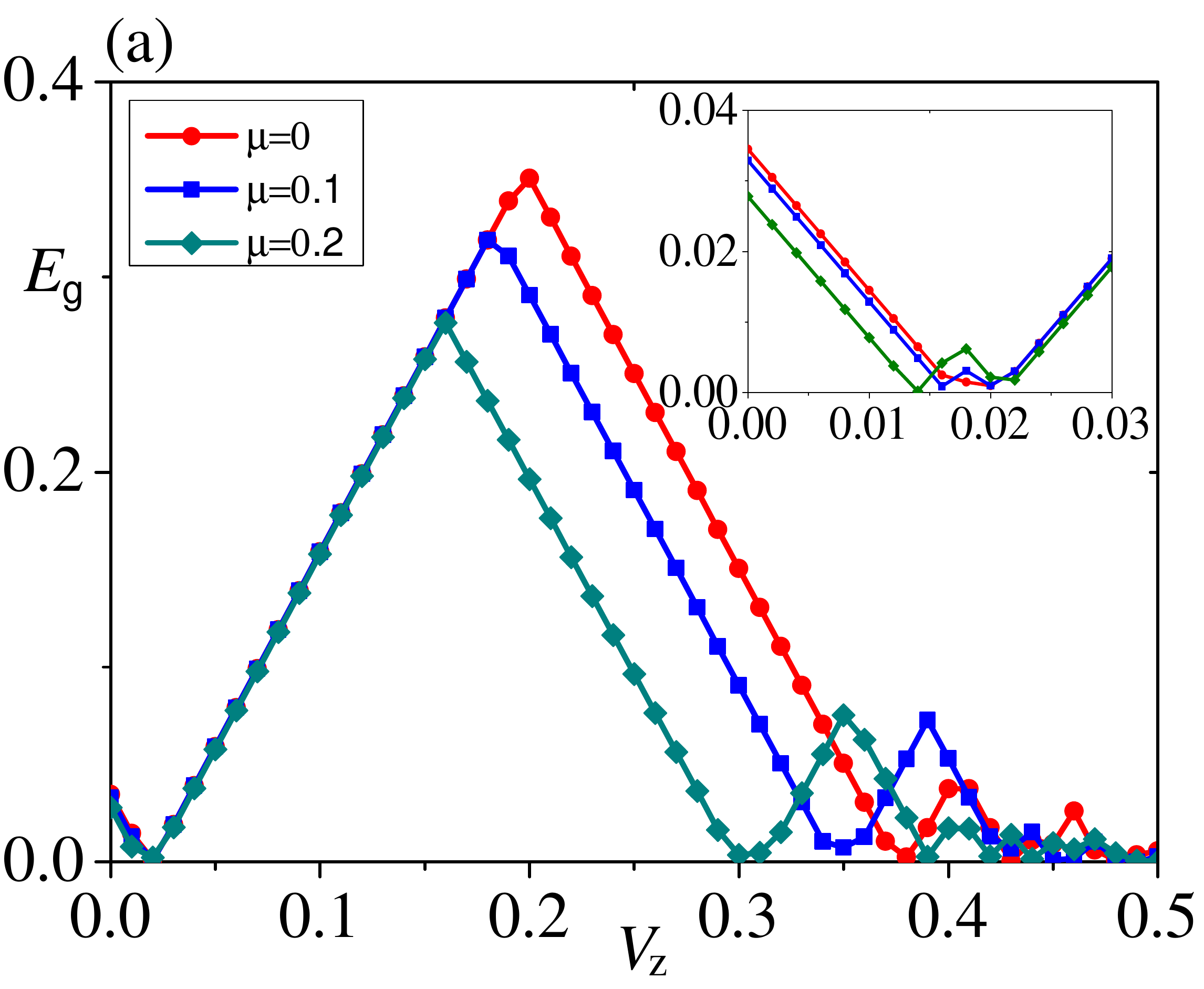}}
\subfigure{\includegraphics[width=4.25cm, height=3.5cm]{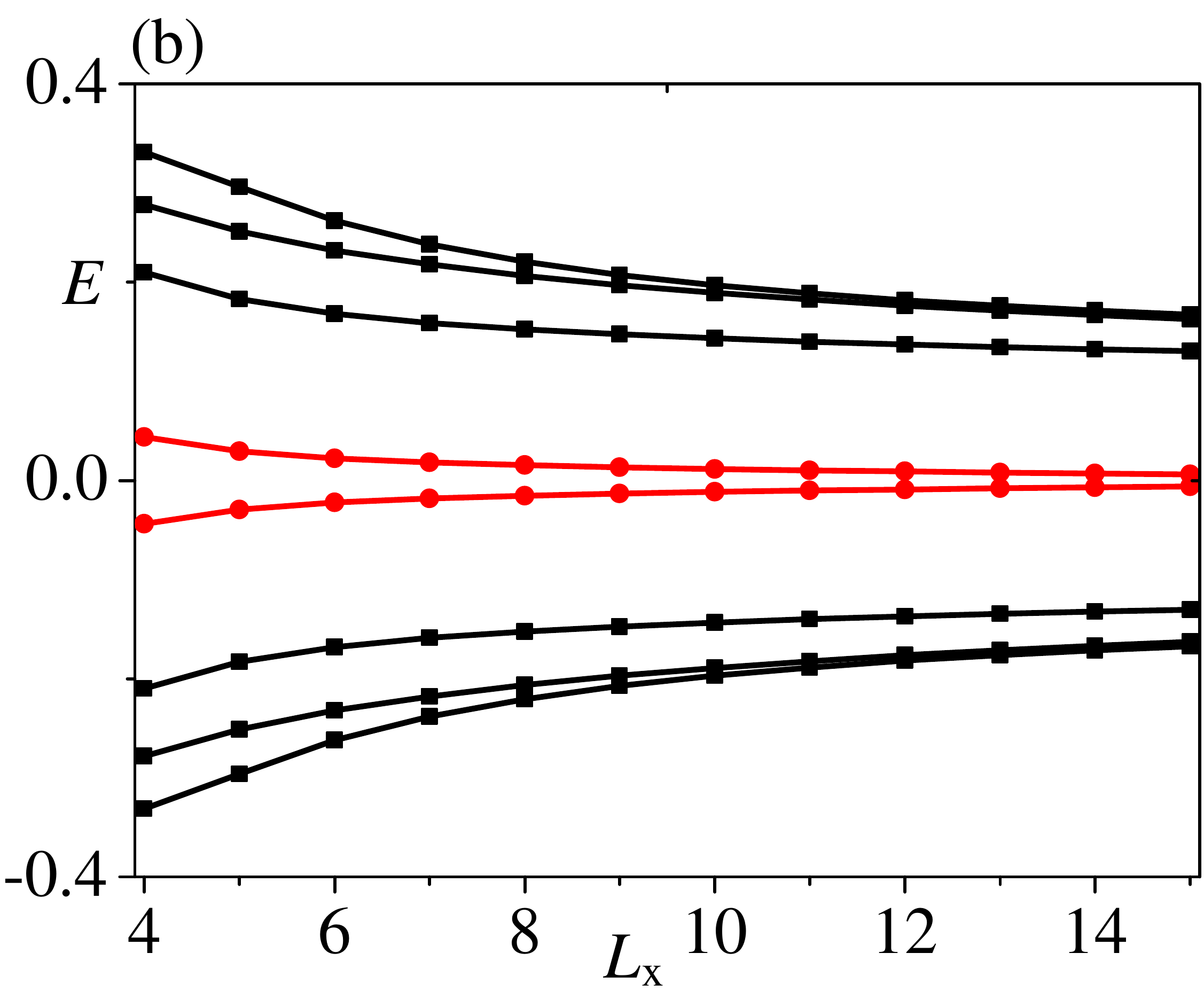}}
\subfigure{\includegraphics[width=4.25cm, height=3.5cm]{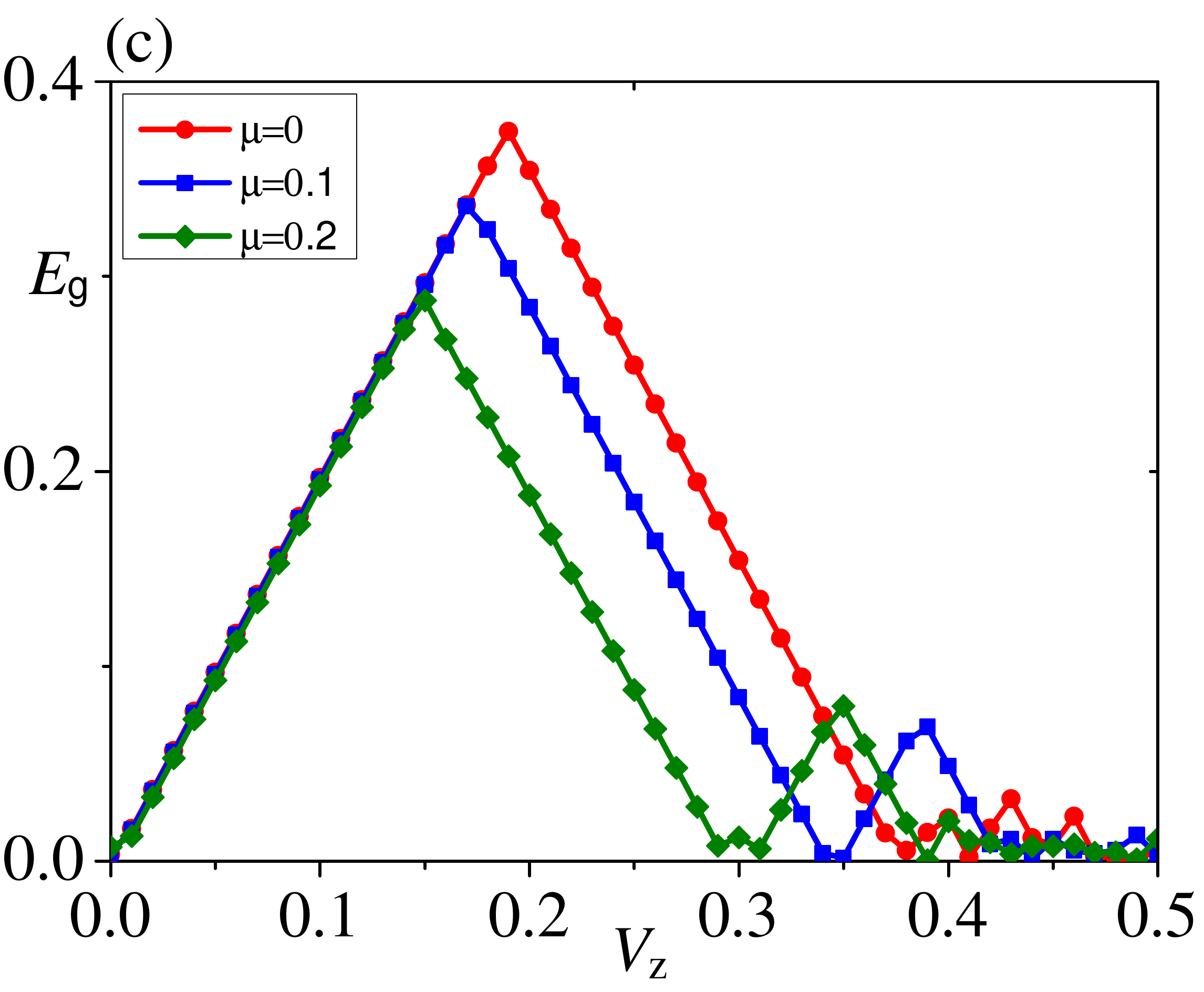}}
\subfigure{\includegraphics[width=4.25cm, height=3.5cm]{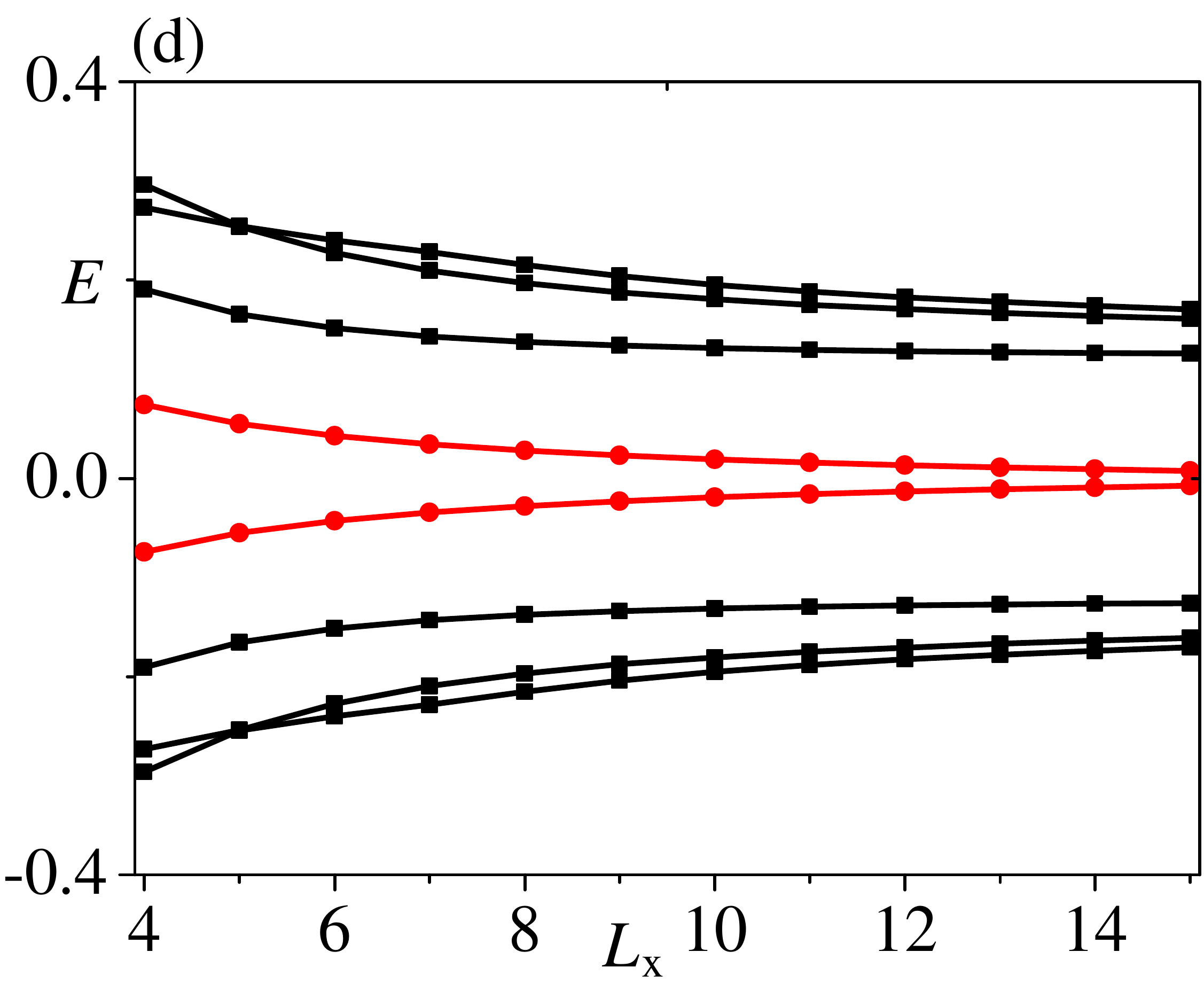}}
\caption{(a)(c) The evolution of energy gap at $k_{x}=0$ with respect to Zeeman field,
the inset in (a) is an enlarged view of the low-field regime. (b)(d) The
energy spectra for a cubic sample with open boundary condition in all three directions (only
the eight energies closet to zero are shown).
Common parameters are
$m=2.5$, $t=t_{3}=t'=1$, $\delta=0$, $\Delta_{0}=0.5$, $\xi=4$.
$L_{y}=L_{z}=20$
in (a)(c), $L_{y}=L_{z}=8$ in (b)(d), $t_{3}'=1$ in (a)(b), $t_{3}'=0$ in (c)(d),
$\mu=0$ and $V_{z}=0.2$ in (b)(d).}  \label{Zeeman}
\end{figure}

Keeping the finite value of $t_{3}'$, Fig.\ref{Zeeman}(a) shows how the
energy gap of the vortex line at $k_{x}=0$ (denoted by $E_{g}$) evolves with
the Zeeman field for $\mu=0, 0.1, 0.2$. It is readily found that
$E_{g}$ first decreases and gets closed twice at two very small and close
$V_{z}$ (see the inset of Fig.\ref{Zeeman}(a)) and then
increases linearly to values much greater than the zero-field value.
Here the double closures of $E_{g}$ are simply because the lowest excitation
spectra of the vortex line are nearly doubly degenerate (see Fig.\ref{vortex}).
Since $\nu$ changes its value for each closure and reopening of $E_{g}$,
$\nu$ returns its zero-field value after entering the gap-linear-increase regime.
Our numerical calculations of $\nu$
confirm the above arguments.
This fact reveals that the Zeeman field can greatly enhance the
topological gap protecting MZMs in the weakly-doped regime.
 To further support this, we diagonalize
the Hamiltonian for a cubic geometry with open boundary condition
in all three directions.  In Fig.\ref{Zeeman}(b), we present the eight energies closest
to $E=0$ for $\mu=0$ and $V_{z}=0.2$ ($\nu=-1$ for this case),
their scalings with the system size clearly reveal
the existence of two MZMs.

Fig.\ref{Zeeman}(c) shows the $E_{g}$-$V_{z}$ dependence in the limit
$t_{3}'=0$. In comparison to Fig.\ref{Zeeman}(a), a qualitative difference
appears in the low-field regime. In Fig.\ref{Zeeman}(c), the energy gap is vanishingly small at
$V_{z}=0$,  indicating the crucial role of the $t_{3}'$ terms in the realization of robust MZMs
when the Zeeman field is negligible. Remarkably, away from the low-field regime,
Figs.\ref{Zeeman}(c) and (a)  show little difference,
revealing the important fact
that vortex-end MZMs can also be realized in Dirac semimetals with only Fermi arcs
(see Fig.\ref{Zeeman}(d)). In sharp contrast, we find that the Zeeman field monotonically  reduces
the topological gap of the vortex lines in superconducting TIs (see details in SM\cite{supplemental}).

\begin{figure}
\subfigure{\includegraphics[width=4.25cm, height=3.5cm]{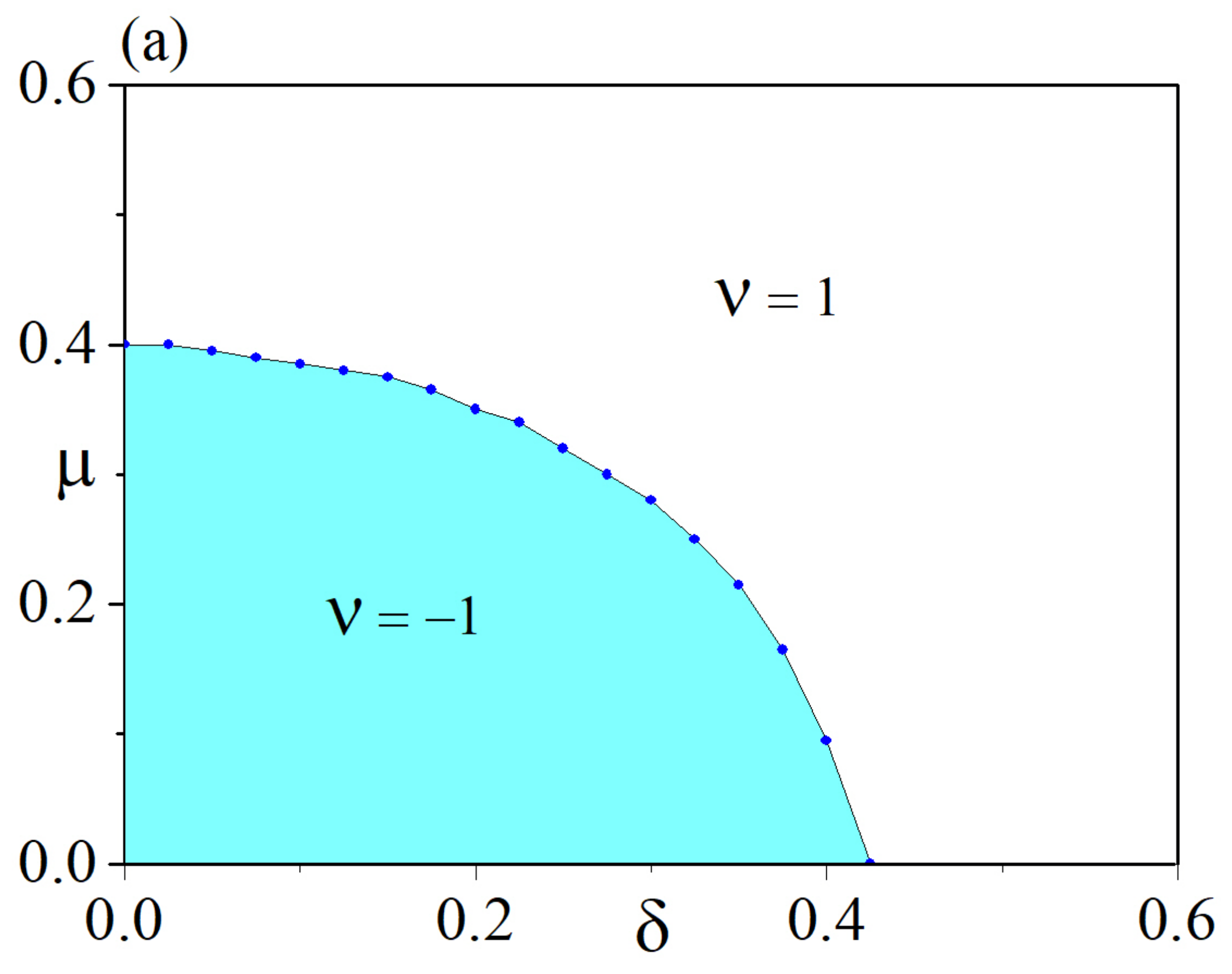}}
\subfigure{\includegraphics[width=4.25cm, height=3.5cm]{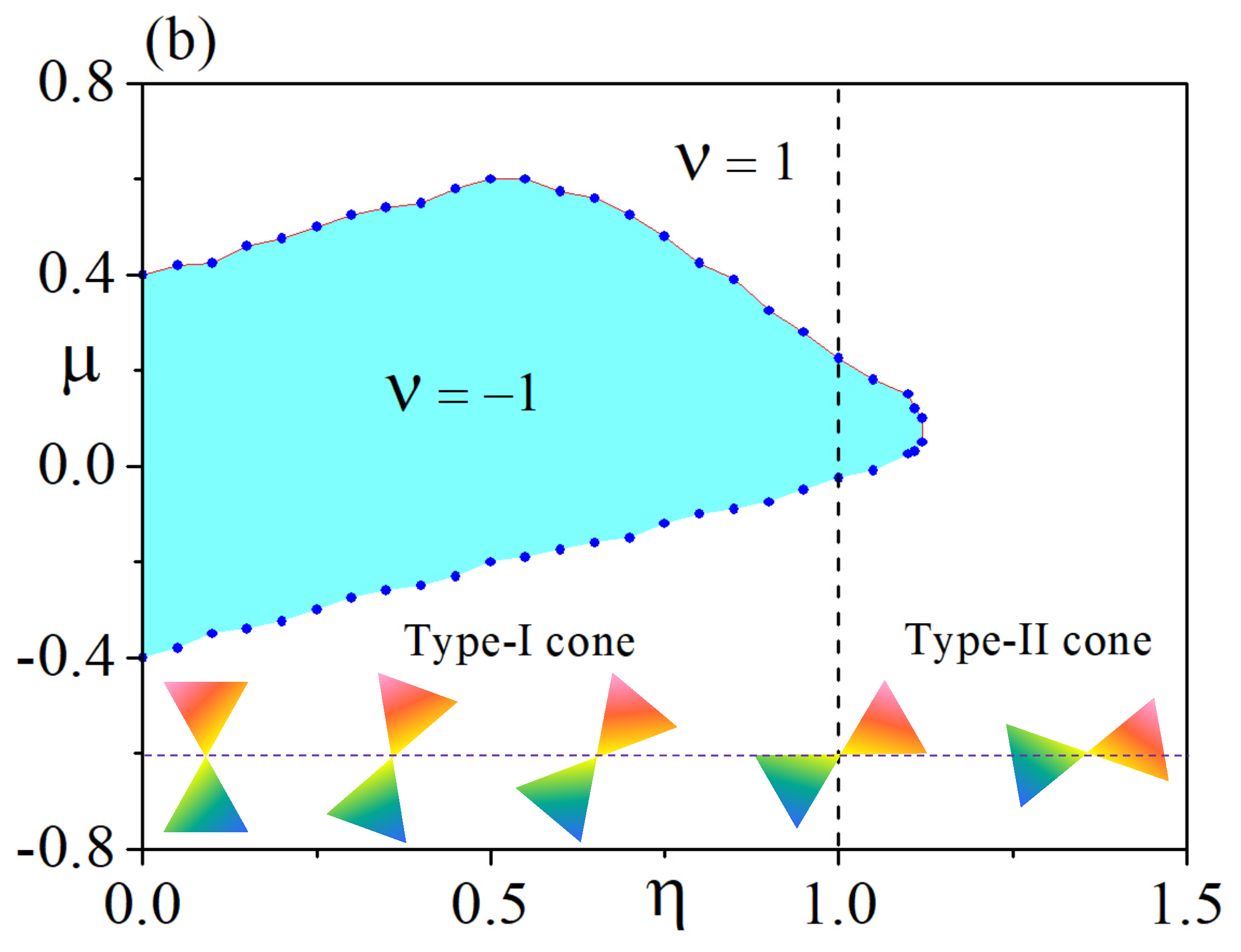}}
\caption{ (a) The topological phase diagram
for the superconducting Weyl semimetal.  (b) The effect of tilting of Dirac cones, characterized by the $\eta$ parameter, on the topological regime of the vortex line. Common parameters in (a) and (b) are $m=2.5$, $t=t_{3}=t'=t_{3}'=1$, $\Delta_{0}=0.5$, $\xi=4$, $L_{y}=L_{z}=20$. }  \label{tilt}
\end{figure}

As the stability of MZMs does not need any crystal symmetry, it is obvious that
vortex-end MZMs can also be realized when $\delta$ is finite but small.
In Fig.\ref{tilt}(a), we present the impact of $\delta$ on $\mu_{c}$. For simplicity,
we only show the result for finite $t_{3}'$ and $V_{z}=0$\cite{supplemental}.
The phase diagram suggests that increasing $\delta$, which increases the separation between
Weyl points, decreases the topological regime but in a considerably smooth way, indicating
that vortex-end MZMs can also be realized in many Weyl semimetals when they become
superconducting.

{\it The effect of tilting.---} Dirac and Weyl semimetals have commonly been divided into two classes, i.e. type-I and type-II~\cite{soluyanov2015type}, in accordance with
the extent of tilting of Dirac and Weyl cones. To capture the effect of tilting, here we let the so far ignored $\varepsilon(\bk)$ term in Hamiltonian (\ref{normal}) take the form $\varepsilon(\bk)=\eta t_{3}(\cos k_{z}-\cos k_{0})$, where $\pm k_{0}$ are  the coordinates of Dirac or Weyl points in the $k_{z}$ direction and $\eta$ characterizes the extent of tilting, with
$\eta<1$ and $\eta>1$ corresponding respectively to the type-I and type-II phase.

Since the tilting effects for the Dirac semimetal and the Weyl semimetal are found to share qualitatively the same picture, here we only present the results for the former and present the results for the latter in the SM\cite{supplemental}. As shown in Fig.\ref{tilt}(b),  the topological regime barely changes for $\eta<0.5$ but it  shrinks as $\eta$ increases further. After entering into the type-II phase, it is found to vanish quickly, suggesting that type-II Dirac semimetals are less favored than type-I ones
for the realization of vortex-end MZMs.

{\it Experimental consideration.---} As  superconductivity has already been observed in
quite a few TRI Dirac and Weyl semimetals~\cite{Hosur2014,Kobayashi2015Dirac,Hashimoto2016DSM,Chen2016,Butch2011,
pan2015pressure,kang2015superconductivity,wang2016observation,aggarwal2016unconventional,wang2017tip,
Bachmanne2017swsm,qi2016superconductivity,Noh2017DSM,Leng2017DSM,Yu2018sdsm,Wang2018sdsm,huang2019proximity},
our predictions are immediately testable in experiment. A very promising class of platforms are semimetal/superconductor hybrid structures, like the experimentally studied
Cd$_{3}$As$_{2}$/Nb\cite{huang2019proximity}. As is known,
Cd$_{3}$As$_{2}$ is an ideal Dirac semimetal with the Fermi energy very close to the two Dirac points\cite{wang2013three},
therefore, when it becomes superconducting due to the proximity effect of the $s$-wave
superconductor Nb,  the vortex lines terminating at the surfaces with Fermi arcs
are expected to host MZMs at their ends. Furthermore,
Dirac points close to Fermi energy  have also been observed in several iron-based superconductors\cite{zhang2019multiple}, e.g., FeTe$_{x}$Se$_{1-x}$ and LiFe$_{1-x}$CoAs.
Our findings can also be tested in these intrinsic superconductors by tuning
the doping level to be closer to the Dirac points.

{\it Conclusion.---} We have revealed that
MZMs can be realized at the ends of a $\pi$-flux vortex line in superconducting
TRI  Dirac and Weyl semimetals
if the vortex line is terminated at the surfaces with
Fermi arcs or loops and the doping is lower than a critical level,
and remarkably,  the Zeeman field can profoundly enhance the stability
of MZMs in these systems. We have also shown that
type-I Dirac and Weyl semimetals are in general more favorable
than type-II ones for the realization of MZMs.
Our findings offer a new perspective on the potential applications of topological
semimetals.

{\it Acknowlegements.---}
Z.Y. would like to thank the hospitality of
Shenzhen Institute for Quantum Science and Engineering at Southern University of Science and Technology
and Peng Cheng Laboratory, where this work was started. Z.Y. acknowledges
the support by the Startup Grant (No. 74130-18841219) and
the National Science Foundation of China (Grant No.11904417). Z.W.
acknowledges the support by the National Science Foundation of China (Grant No.11904417)
and the Key-Area Research and Development Program of GuangDong Province (Grant No.2019B030330001). W.H.  is supported by the National Science Foundation of China (Grant No. 11904155), and by the Guangdong Provincial Key Laboratory (Grant No.2019B121203002).

\bibliography{dirac}

\end{document}